\documentclass[aps,prl,twocolumn,showpacs]{revtex4}
\usepackage{amssymb}
\usepackage{amsmath, amsthm}
\newcommand{\be}{\begin{equation}}
\newcommand{\ee}{\end{equation}}

\begin{document}


\title{The Lagrangian description of perfect fluids and modified 
gravity with an extra force} 


\author{Valerio Faraoni}
\email[]{vfaraoni@ubishops.ca}
\affiliation{Physics Department, Bishop's University\\
Sherbrooke, Qu\'ebec, Canada J1M~1Z7
}

\begin{abstract} 
We revisit the issue of the correct Lagrangian  description of a 
perfect fluid (${\cal L}_1=P$ versus   ${\cal L}_2=-\rho$) in 
relation with  modified gravity theories in  which 
galactic luminous 
matter couples nonminimally to the Ricci scalar. These 
Lagrangians are only equivalent when the fluid couples minimally 
to gravity and not otherwise; in the presence of nonminimal 
coupling they give rise to two distinct theories with different 
predictions.

\end{abstract}


\maketitle

\section{Introduction}
\setcounter{equation}{0}

Attempts to explain the present acceleration of the universe 
discovered with type Ia supernovae \cite{SN} without invoking 
a mysterious and exotic dark energy \cite{Linder} have lead to 
modifying gravity at the largest scales in the so-called 
$f(R)$ gravity (\cite{CCT}, see \cite{review} for a 
comprehensive review and  \cite{otherreviews} for brief 
introductions). $f(R)$ theories of gravity have been employed 
also 
in attempts to explain dark matter in galaxies and clusters 
\cite{f(R)darkmatter}. A recent proposal ny Bertolami, 
B\"{o}hmer, Harko, and Lobo (BBHL) \cite{BBHL} contemplates 
the possibility of coupling matter explicitly to the Ricci 
curvature, as described by the action \footnote{This action was 
introduced in \cite{Allemandietal2005}.}
\be\label{BBHLaction}
S_{BBHL}=\int d^4x \, \sqrt{-g} \, \left\{ f_1(R) +\left[ 
1+ \lambda f_2(R) \right] 
{\cal L}^{(m)} \right\} \,,
\ee
where $R$ is the Ricci curvature of spacetime, $f_{1,2}(R)$ are 
functions of $R$, $\lambda$ is  a small parameter, and ${\cal 
L}^{(m)}$ is the matter Lagrangian density (see \cite{seealso} 
for  similar theories and \cite{nonminimalem} for the special 
case of a gauge field coupling explicitly to the 
curvature). The extra coupling leads to the non-conservation of 
the matter stress-energy tensor $T_{ab}^{(m)}$, according to
\be
\nabla^b T_{ab}^{(m)}=\frac{\lambda f_2'(R)}{1+\lambda f_2(R)} 
\left[ g_{ab} {\cal L}^{(m)}-T_{ab}^{(m)} \right] \nabla^b R \,,
\ee
where a prime denotes differentiation with respect to $R$. 
In  \cite{BBHL} BBHL found that an extra force, not 
appearing in the minimally 
coupled versions of $f(R)$ gravity, acts upon particles and could 
effectively replace dark matter. This intriguing possibility has 
motivated further work \cite{BBHL2, BLP, othersonBBHL}.

In \cite{BBHL} a dust fluid model is assumed  
to describe ordinary (luminous) matter in galaxies and the extra 
force
\be
f^{a}=\left( -\frac{\lambda f_2'}{1+\lambda f_2}\, \nabla_b 
R+\frac{\nabla_b P}{P+\rho} \right)h^{ab}
\ee
was derived (here $u^c$ is the fluid four-velocity and 
$h_{ab} \equiv g_{ab}+u_a u_b$).  In \cite{SotiriouFaraoni} it 
was noted 
that, adopting the  standard Lagrangian density ${\cal L}_1=P$ 
for a perfect fluid, this  
extra force vanishes for dust, which has equation of state $P=0$ 
adequately reproducing the non-relativistic motions of stars, 
making again dark matter a necessary ingredient to explain the 
galactic rotation curves.  
Subsequently Bertolami and collaborators pointed out that an 
equivalent Lagrangian density for  a perfect fluid is $ {\cal 
L}_2=-\rho $, which is obtained by adding surface terms to the 
action $S_1=\int d^4x \, \sqrt{-g} \, {\cal L}_1$  \cite{BLP} 
(see \cite{SchutzBrown} for detailed studies of  the 
Lagrangian formalism for   perfect 
fluids \footnote{The Lagrangian density ${\cal 
L}_2=-\rho$ is used less frequently than ${\cal L}_1=P$; it 
appears,  {\em e.g.}, in Ref.~\cite{HawkingEllis}.}). From 
\cite{BLP} it would appear 
that the extra force that has the potential to replace dark 
matter is present or absent according to the choice that is made 
between  the two  supposedly equivalent Lagrangians. This shows 
that, 
clearly, the two Lagrangian densities ${\cal L}_1=P$ and ${\cal 
L}_2=-\rho$ cannot be equivalent, and the recent literature has 
discussed this issue in terms of the problem of ``which 
Lagrangian density   correctly describes a perfect fluid''. Here  
we show that, posed in these terms, this problem is 
meaningless. In fact, for a perfect fluid that does not couple 
explicitly to the curvature ({\em i.e.}, for $\lambda =0$), the 
two Lagrangian densities ${\cal L}_1=P$ and ${\cal L}_2=-\rho$ 
are perfectly equivalent, as discussed extensively in 
Refs.~\cite{SchutzBrown} and remarked in \cite{BLP}. However, 
there is little doubt that for a coupled perfect fluid 
($\lambda \neq 0$) the two Lagrangians are inequivalent. This 
fact is a manifestation of a more general situation: if a 
Lagrangian system consists of two subsystems and there are two 
Lagrangians for one of these subsystems, which provide  
equivalent descriptions for that subsystems {\em when it is 
isolated}, the two Lagrangians cease to be equivalent {\em when 
the subsystem couples to the rest of the system}. An elementary 
example is provided in the next section. We follow the notations of 
Ref.~\cite{Wald}.

\section{Two coupled oscillators} 

Point particle mechanics provides a very simple example of this 
situation. Considers two coupled one-dimensional oscillators  
 which, in isolation, are described by the Lagrangians
\begin{eqnarray}
L_1 \left( q_1, \dot{q}_1 \right) & = & \frac{1}{2} \, \left( 
\dot{q}_1  \right)^2  -\frac{K_1}{2} \, q_1^2 \,,\\
&&\nonumber\\
L_2 \left( q_2, \dot{q}_2 \right) & = & \frac{1}{2} \, \left( 
\dot{q}_2 \right)^2 
-\frac{K_2}{2}\, q_2^2 \,,
\end{eqnarray}
where $q_{1,2}$ are the respective Lagrangian coordinates.  Now 
couple the two oscillators via the term $L_{12} = -\lambda q_1 
L_2  $ in the total Lagrangian
\begin{eqnarray}
&& L \left( q_1, \dot{q}_1, q_2, \dot{q}_2 \right) = 
L_1+L_2+L_{12}=
\frac{1}{2} 
\left[ \left( \dot{q}_1 \right)^2 + \left( \dot{q}_2  \right)^2 
\right] \nonumber\\
&&\nonumber\\
&&  -\frac{K_1}{2}\, q_1^2 -\frac{K_2}{2}\, q_2^2 
-\lambda q_1 \left[
\frac{1}{2} \, \left( \dot{q}_2 
\right)^2 
-\frac{K_2}{2}\, q_2^2 \right] \;.
\end{eqnarray}
The Euler-Lagrange equations 
\be
\frac{d}{dt} \left( \frac{\partial L}{\partial \dot{q}_i} 
\right)-\frac{\partial L}{\partial q_i}=0 \;\;\;\;\;\;\; (i=1,2)
\ee
yield the equations of motion 
\begin{eqnarray}
&& \ddot{q}_1+K_1 q_1 +\frac{\lambda}{2} \left[ \left( 
\dot{q}_2\right)^2 -K_2 q_2^2 \right]=0 
\,,\label{firstequation}\\
&&\nonumber\\
&& \ddot{q}_2 \left( 1-\lambda q_1 \right)  -\lambda\dot{q}_1 
\dot{q}_2 + K_2 \left( 1-\lambda q_1 \right) q_2=0 \,.
\end{eqnarray}
Now consider the new Lagrangian $L_2+C$ (where $C$ is a 
nonzero constant) for the subsystem~2: when this subsystem is 
isolated, this new Lagrangian  is 
trivially equivalent to $L_2$. However, see what happens when 
$L_2$ is replaced by $L_2+C$ in the description of the {\em 
coupled} subsystem; the equations of motion become then 
\begin{eqnarray}
&& \ddot{q}_1 + K_1 q_1 +\frac{\lambda}{2} \left[ \left( 
\dot{q}_2\right)^2 -K_2 q_2^2 \right] +\lambda C =0 
\,,\label{firsteqprime}\\ 
&&\nonumber\\
&& \ddot{q}_2 \left( 1-\lambda q_1 \right)  -\lambda\dot{q}_1 
\dot{q}_2 +K_2 \left( 1-\lambda q_1 \right) q_2=0 \,.
\end{eqnarray}
Eq.~(\ref{firsteqprime}) does not coincide with 
(\ref{firstequation}).

To give another, less trivial, example assume that $L_2$ is 
replaced by the equivalent 
Lagrangian (when the subsystem~2 is isolated) 
$L_2+d F /dt$, 
where $ F(q_2, t)$ is a function of the coordinate $q_2$ and 
the time $t$. Then, the  total Lagrangian of the two coupled 
subsytems becomes 
\be
L^*=L_1+L_2 +\frac{dF}{dt}-\lambda q_1 L_2 -\lambda q_1 
\frac{dF}{dt} \,;
\ee
the change,  which leaves  the equation of motion for the 
oscillator~2 unaffected when the latter is isolated, now affects 
the way this oscillator interacts with the first one. The new 
equations of motion are 
\begin{eqnarray}
&& \ddot{q}_1+K_1 q_1 +\frac{\lambda}{2} \left[ 
\left( \dot{q}_2 \right)^2 -K_2 
q_2^2 \right] +\lambda \, \frac{\partial F}{\partial q_2}\, 
\dot{q}_2 +\lambda \, \frac{ \partial F}{\partial t}=0 
\,,\nonumber\\
&& \label{abra} \\
&& \ddot{q}_2\left( 1-\lambda q_1 \right) -\lambda \dot{q}_1 
\left( \dot{q}_2 +\frac{\partial F}{\partial q_2} \right) 
 + K_2\left( 
1-\lambda q_1 \right) q_2  =0 \,.\nonumber\\
&&\label{cadabra}
\end{eqnarray}

\section{Discussion and conclusions}

Another argument advocated in \cite{BLP}  as evidence for the 
extra force in galaxies needs to be addressed: in  
Ref.~\cite{PuetzfeldObukhov}, Puetzfeld and Obukhov derive the 
equations of motion of a test body described by a Lagrangian 
density ${\cal L}$ and  an energy-momentum tensor $T_{ab}^{(m)}$ 
in the 
modified gravity theories introduced in \cite{BBHL}, using  
a multipole method. They show that, in general, extra forces 
occur 
even on single-pole particles, as described by  eq.~(32) 
of \cite{PuetzfeldObukhov},
\be
\frac{D}{D \tau} \left[ mu^i \left( 1+\lambda f_2 \right) \right] 
=N^{ib} \lambda \nabla_b f_2 \,.
\ee
Here $\tau$ is the proper time along the worldline of a particle 
with four-tangent $u^a$ and mass $m$, while  
\be
N^{ab} \equiv u^0 \Xi^{ab} = u^0 \int_{\Sigma(t)}  d^3 \vec{x} 
\, \tilde{\Xi}^{ab} \,,
\ee
where a tilde denotes the density of the corresponding 
quantity and $\Xi^{ab} \equiv {\cal L} g^{ab}$ 
\cite{PuetzfeldObukhov}. In spite of being a significant piece 
of work describing the motion of extended objects, the analysis  
of  \cite{PuetzfeldObukhov} does not solve the issue of 
which Lagrangian (${\cal L}_1$ or ${\cal L}_2$) is appropriate to 
describe a perfect fluid: it merely says what are the equations 
of motion once a Lagrangian is assumed. If we assume the 
Lagrangian ${\cal L}_1=P$, then $\Xi^{ab}=Pg^{ab}$ and, 
consequently, $N^{ab}$, vanish for a dust with $P=0$. If 
instead the Lagrangian ${\cal L}_2=-\rho$ is assumed, $N^{ab}$ 
does not vanish and there are extra forces even on the particles 
of a  dust fluid. Hence, the work of \cite{PuetzfeldObukhov} 
des not resolve the issue of whether ${\cal L}_1$ or ${\cal L}_2$ 
is appropriate to describe a perfect fluid nonminimally coupled 
to gravity.

We note that the case of the cosmological constant, regarded as 
an effective form of matter,  is very special: if one considers a 
cosmological constant as a perfect fluid with Lagrangian ${\cal 
L}=-\Lambda$, there is no difference between the two Lagrangians 
${\cal L}_1=P$ and ${\cal L}_2=-\rho$ because of the peculiar 
equation of state $P=-\rho$ of the effective fluid, and there is 
no extra force on the cosmological constant ``fluid''. This 
coincidence is, of course, consistent with the fact that one can 
also regard the cosmological constant term as a purely 
geometrical term which is part of the first function $f_1(R)$ in 
the action~(\ref{BBHLaction}) (for example, $f_1(R)=R-\Lambda$).

To summarize our conclusions, there is little doubt that the  two 
Lagrangian densities ${\cal 
L}_1=P$ and ${\cal L}_2=-\rho$ are equivalent for the description 
of a perfect fluid which is not coupled directly to gravity, as 
shown in Refs.~\cite{SchutzBrown}. However, as soon as this fluid 
is coupled explicitly to gravity as in eq.~(\ref{BBHLaction}), 
the two Lagrangian densities cease to be equivalent. It is not 
clear that one should be physically preferred with respect to 
the other: simply, they give rise to two inequivalent theories of 
gravity and matter, which are both correct. Which one should be 
chosen must be decided by independent arguments. It is a fact 
that by choosing ${\cal L}=P$ there is no extra force on a dust 
fluid and 
it is equally undeniable that by choosing ${\cal L}_2=-\rho$ 
there will be such a force, which may ultimately provide an 
alternative to dark matter. We are not able to provide 
independent arguments in favour of one choice or the other: the 
contribution of the present note merely consists in showing that, 
in spite of the appearance,  there is no  contradiction between 
the 
results of   \cite{SotiriouFaraoni} and 
those of \cite{BLP}.

\begin{acknowledgments}

The author thanks Dirk Puetzfeld for bringing 
Ref.~\cite{PuetzfeldObukhov} to his attention. This work is 
supported by the  Natural Sciences and  Engineering Research 
Council of Canada ({\em NSERC}). 
\end{acknowledgments}


\begin{thebibliography}{99}

\bibitem{SN} A.G. Riess {\em et al.}, {\em Astron. J.}
{\bf 116}, 1009 (1998);  {\em Astron. J.} {\bf 118}, 2668 (1999);
{\em Astrophys. J.} {\bf 560}, 49 (2001);
{\em Astrophys. J.} {\bf 607}, 665 (2004);
S. Perlmutter {\em et al.}, {\em Nature} {\bf 391}, 51 (1998);
{\em Astrophys. J.} {\bf 517}, 565 (1999);
 J.L. Tonry {\em et al.}, {\em Astrophys. J.} {\bf 594},
1 (2003);
R. Knop {\em et al.}, {\em Astrophys. J.} {\bf 598}, 102 (2003);
B. Barris {\em et al.}, {\em Astrophys. J.} {\bf 602}, 571 
(2004).

\bibitem{Linder} E.V. Linder, {\em Am. J. Phys.} {\bf
76}, 197 (2008).

\bibitem{CCT} S. Capozziello, S. Carloni, and  
A. Troisi, {\em Recent Res. Dev. Astron. Astrophys.} {\bf 
1}, 625 (2003); 
S.M. Carroll, V.  Duvvuri, M. Trodden, and  
M.S. Turner, {\em Phys. Rev. D} {\bf 70}, 043528 (2004);
S. Nojiri and S.D. Odintsov, {\em Phys. Rev. D} {\bf 68}, 123512 
(2003).



\bibitem{review} T.P. Sotiriou and V. Faraoni,
arXiv:0805.1726, to appear in {\em Rev. Mod. Phys.}

\bibitem{otherreviews}
S. Nojiri and S.D. Odintsov, {\em Int. J. Geom. Meth. Mod. Phys.}
{\bf 4}, 115 (2007); N. Straumann, arXiv:0809.5148;
H.-J. Schmidt,  {\em Int. J. Geom. Meth. Phys.} {\bf 4}, 209 
(2007); V. Faraoni, arXiv:0810.2602; T.P.
Sotiriou, arXiv:0805.1726; S. Capozziello, M. De Laurentis, and 
V. Faraoni, arXiv:0909.4672. 

\bibitem{f(R)darkmatter} S.  Capozziello, E. De Filippis, V. 
Salzano, {\it Mon. Not. R. Astr. Soc.}  {\bf 394},   947 (2009); 
S.  Capozziello, V.F. Cardone, and A. Troisi, {\it Mon. Not. R. 
Astr. Soc.}  {\bf 375},   1423 (2007); C.G. Boehmer, T. Harko, 
and F.S. Lobo, {\em Astropart. Phys.} {\bf 29}, 386 (2008); {\em 
J. Cosmol. Astropart. Phys.} {\bf 0803}, 024 (2008). 

\bibitem{BBHL} O. Bertolami , C.G. B\"{o}hmer, T. Harko, and 
F.S.N. Lobo, {\em Phys. Rev. D} {\bf 75}, 104016 (2007).

\bibitem{Allemandietal2005} G. Allemandi, A. Borowiec, M. 
Francaviglia, and S.D. Odintsov, {\em Phys. Rev. D} {\bf 72}, 
063505 (2005).

\bibitem{seealso} 
S. Nojiri and S.D. Odintsov,  {\em Phys. 
Lett. B} {\bf 599}, 137 (2004); 
S. Nojiri, S.D. Odintsov, and P. Tretyakov, 
{\em Progr. Theor. Phys. (Suppl.)} {\bf 172}, 81 (2008);
S. Mukohyama  and L. Randall, {\em 
Phys. Rev. Lett.} {\bf 92}, 211302 (2004); 
S. Nojiri and S.D. Odintsov, {\em 
Proc.  Sci. WC} 024 (2004); T. Inagaki, S. Nojiri, and 
S.D. Odintsov,  {\em JCAP} {\bf 0506} 010 (2005).


\bibitem{nonminimalem}  M. Novello  and J.M. Salim,  {\em 
Phys. Rev. D} {\bf 20}, 377 (1979);  
M. Novello and H. Heintzmann, {\em Gen. Rel. Grav.} {\bf 
16}, 535 (1984); 
M.S. Turner  and  L.M. Widrow, {\em Phys. Rev. D} 
{\bf 37},  2743 (1988);
M. Novello, V.M.C.  Pereira, and N. Pinto-Neto, {\em Int. 
J. Mod. Phys. D} {\bf 4}, 673 (1995);
R. Lafrance  and R.C. Myers, {\em Phys. Rev. D} 
{\bf 51}, 2584 (1995); K. Bamba  and S.D. Odintsov, 
{\em JCAP} 0804:024 (2008); K. Bamba, S. Nojiri, and S.D. 
Odintsov, {\em Phys. Rev. D} {\bf 77}, 123532 (2008).

\bibitem{BBHL2} O. Bertolami and J. P\'{a}ramos,  {\em 
Phys. Rev.  D} {\bf 77}, 084018 (2008).

\bibitem{BLP} O. Bertolami, F.S.N. Lobo, and J. P\'{a}ramos, {\em 
Phys. Rev. D} {\bf 78}, 064036 (2008); O. Bertolami, J. 
P\'{a}ramos, T. Harko, and F.S.N. Lobo, arXiv:0811.2876.

\bibitem{othersonBBHL} 
V. Faraoni, {\em Phys. Rev. D} {\bf 76}, 127501 (2007); O. 
Bertolami and J. P\'{a}ramos, {\em Class. Quantum Grav.} {\bf 
25},  245017 (2008); T.P. Sotiriou, {\em Phys. Lett.  B} {\bf 
664},  225 (2008); O. Bertolami and M. Carvalho Sequeira, {\em 
Phys. Rev. D} {\bf 79}, 104010 (2009); O. Bertolami and J. 
P\'{a}ramos, arXiv:0906.4757; O. Bertolami and M. Carvalho 
Sequeira,  arXiv:0910.3876.

\bibitem{SotiriouFaraoni} T.P. Sotiriou and V. Faraoni, {\em 
Class. Quantum Grav.} {\bf 25}, 205002 (2008).

\bibitem{SchutzBrown} R.L.  Seliger and G.B.  Whitham,  
{\em Proc. R. Soc. (London)} {\bf A305}, 1 (1968); 
B. Schutz 1970 {\em Phys. Rev. D} {\bf 2}, 2762 (1970); J.D. 
Brown, {\em Class. Quantum Grav.} {\bf 10}, 1579 (1993).

\bibitem{Wald} R.M. Wald, {\em General Relativity} (Chicago 
University Press, Chicago, 1984).

\bibitem{PuetzfeldObukhov} D. Puetzfeld and Y.N. Obukhov, {\em 
Phys. Rev. D} {\bf 78}, 121501(R) (2008).

\bibitem{HawkingEllis} S.W. Hawking  and G.F.R. Ellis,  
{\em The Large Scale Structure of Space-Time} (Cambridge 
University Press, Cambridge, 1973).

\end{thebibliography}

\end{document}